\documentclass[useAMS,usenatbib]{mn2e}
\usepackage{graphicx}
\usepackage{caption}
\usepackage{epsfig}
\usepackage{amsfonts}
\usepackage{natbib}

\newcommand{\fe}[0]{[{\rm Fe/H}]}

\newcommand{\kpc}[0]{{\rm \, kpc}}

\newcommand{\vlos}[0]{v_{\rm los}}

\def\mag{{\rm \, mag}}

\def\kms{{{\rm km}s^{-1}}}

\title[A new calibration for the Blue Horizontal Branch]{A new calibration for the Blue Horizontal Branch}

\author[Fermani \& Sch\"onrich]{Francesco Fermani$^{1}$\thanks{E-mail:f.fermani1@physics.ox.ac.uk}, Ralph Sch\"onrich$^{2}$\\
$^{1}$University of Oxford, Rudolf Peierls Centre for Theoretical Physics, Oxford OX1 3NP\\
$^{2}$Hubble Fellow, Department of Astronomy, Ohio State University, Columbus, USA}

\begin{document}

\date{Accepted 2012 December 21. Received 2012 December 20; in original form 2012 September 17}

\pagerange{\pageref{firstpage}--\pageref{lastpage}} \pubyear{2013}


\maketitle

\label{firstpage}

\begin{abstract}

We suggest a simple analytic approximation for magnitudes and hence distances of Blue Horizontal Branch (BHB) stars in Sloan colours. Precedent formulations do not offer a simple closed formula, nor do they cover the full dependences, e.g. on metallicity. Further, using BHB star samples from the Sloan Digital Sky Survey (SDSS) we validate our distance calibration directly on field stars instead of globular clusters and assess the performance of other available distance calibrations. Our method to statistically measure distances is sufficiently accurate to measure the colour and metallicity dependence on our sample and can be applied to other sets of stars or filters.

\end{abstract}

\begin{keywords}

Galaxy: halo  -- stars: horizontal branch -- stars: distances

\end{keywords}

\section{Introduction}
Blue Horizontal Branch (BHB) stars are among the most popular tracers of the stellar halo \citep[e.g.][]{pre91,sir04}. They are bright enough to cover distances larger than $5 \kpc$ and span a relatively narrow range in absolute visible magnitude, if one avoids the blue hook \citep[see][for a definition]{whi98,dcr00,bro01}, where the luminosities of the blue-most BHB stars are greatly diminished. Further they have very blue colours, reducing the risk of being confused with other types of stars and are globally available - in contrast to RR Lyrae stars, which, in addition to spectroscopy, require a good coverage in time to resolve their oscillations \citep[][]{Sesar10}.

By far the largest sample of BHB stars comes from the Sloan Digital Sky Survey \citep[SDSS,][]{eis11}. Some previous studies attempted to provide a magnitude calibration for BHB stars in the $ugriz$ filter system of SDSS, based on a combination of theoretical expectations and calibrating on globular clusters \citep[e.g.][]{sir04,x08}. Yet those calibrations have not been provided in a user-friendly form, such that many current studies \citep[e.g.][]{de1, dep10, nie10} still use the approximation of a constant absolute magnitude $M_g=0.7$, originally derived in \cite{yan00} by comparing RR Lyrae and BHB stars in the globular cluster Pal 5.

More importantly, the classic calibrations \citep[like][]{sir04} rely on globular clusters, bearing the usual uncertainties in their distance moduli and reddenings. Further we know that (a) the Horizontal Branch (HB) stars of clusters vary between clusters for reasons that are incompletely understood \citep[e.g. clusters of the same metallicity can either be very red or blue: specifically metal-poor \emph{clusters with high central density present bluer HB morphologies,}][]{sun91,buo93} and (b) there are systematic differences between globular-cluster and field halo stars \citep[e.g. the binary fraction among extreme HB stars is an order of magnitude lower in globular clusters than among field stars,][]{han08}.

Given these problems, it is crucial to create a formula that fulfils the following criteria:
\begin{itemize}
\item{ it covers the principal dependences of the HB magnitudes (i.e. colour and metallicity)}
\item{ it is provided as a simple analytical formula that gives the absolute magnitude from observables;}
\item{ it is validated by/calibrated on field stars.}
\end{itemize}

The aim of this paper is to construct such a calibration in the $ugriz$ filter system of SDSS and assess its performance w.r.t. currently available calibrations by testing it on BHB field stars. At first glance the last point seems out of reach, considering that there are virtually no reliable Hipparcos parallaxes for BHB stars and especially none for stars within the magnitude range of the SDSS. However, the sample size of the SDSS enables us to obtain statistical distance estimates for field BHB stars via the $V$-$W$ correlation method laid out by \cite{SBA}. Even though this exercise is performed on SDSS/SEGUE, our method is general and can be applied to any stellar sample of sufficient size and quality ‪in any filter set‬.

Models of stellar evolution predict the locus of the HB. Hence, our empirical calibration will also provide a test of these models. We find that our calibration agrees quite closely with BASTI isochrones \citep[][]{pie04,pie06}, and makes stars significantly more luminous than Dartmouth isochrones predict \citep[][]{dot07,dot08}.

\section{General method}
We will first identify a general method to assess the accuracy of any distance calibration and then formulate one that fulfils the requirements outlined in the introduction. In principle one could use the distance measurement in the following subsection directly to produce a distance calibration that solely relies on the measured data. As we will see on the BHB star dataset our distance calibration derived from BASTI models \citep[][]{pie04,pie06} is a significant improvement w.r.t. to precedent approaches. Since it is within the measurement errors indistinguishable from the optimum fit to the field star data, we stick with the formula directly derived from theoretical models.

\subsection{How to assess a distance calibration}\label{sec:two}
\cite{SBA}, hereafter SBA12, investigated how distance errors introduce correlations between velocity components. They consider a distance fractional error $f$ such that the measured average distance is $s^\prime=(1+f)s$, where $s$ is the true distance. The observed $(U,V,W)$ velocity components can then be expressed as functions of both the real velocities and of $f$, which introduces correlations between the velocity components. The main method of SBA12 is not applicable for BHB stars, since the (uncertain) proper-motion errors enter the correlation multiplied with the square of the distance. In practice, with current proper motions the method is applicable to samples of stars with $s < 5 \kpc$.

However, we can directly use the linear estimator given in their equations 42-45. This estimator uses the fact that halo stars have a large mean heliocentric azimuthal ($V$) velocity because the halo does not have any significant rotation so we see the full reflex motion of the Sun. Any mean distance error then translates this mean motion into a mean motion in the radial $U$ and vertical $W$ components of heliocentric velocity depending on galactic angles according to the matrix elements provided by SBA12. For the correlation between mean rotational motion and vertical heliocentric mean velocity we have (equation 45 of SBA12):

\begin{equation}
\langle W \rangle + W_\odot = c \cdot T_{WV} + \varepsilon
\end{equation}

\noindent where $c=f \langle V \rangle /(1+f \langle T_{VV} \rangle)$ is the slope of the linear best fit between $\langle W \rangle + W_\odot$, while $T_{VV}=1-\cos^2 b \sin^2 l$ and $T_{WV}=-\frac{1}{2}\sin 2 b \sin l$ are the angular terms through which distance errors introduce correlations between the apparent components of velocity (see Table 1 of SBA12). In particular a correlation between $W$ and $T_{WV}$ implies that the stellar halo is crowding towards the disk on one side of the Galaxy (w.r.t. the Sun-GC line) and dispersing away from the disk on the other side (the ``falling sky " effect, in the nomenclature of SBA12).

In practice, we perform a linear least squares fit of $\langle W \rangle + W_\odot$ versus $T_{WV}$ to estimate the correlation term $c$. Then, we iteratively multiply all distances by correction factors until this correlation on the sky disappears. The fractional average distance error is then just given by the correction factor that makes this correlation zero and the uncertainty on this estimate directly translates into the error bars of $c$. As the method is linear, random proper-motion errors are irrelevant, apart from increasing the error bars by expanding the measured dispersion in $W$.  We are, however vulnerable to systematics in the proper motions. Hence we use the proper-motion correction worked out by \cite{sch12} on the \cite{schn10} quasar sample in Galactic angles $(l, b)$.

\subsubsection{Internal accuracy of the method}

We assess the internal accuracy of the method by testing its ability to recover the mean fractional distance error on a range of control cases. In each of these, we introduce an input bias in the distances and proper motions of the stars in the X11 sample so that the average distance offset is $f_{\rm in}$:
\begin{equation}\label{av_cnd}
\left\langle \frac{d_i-\tilde{d_i}}{d_i} \right\rangle=f_{\rm in} \quad $,$
\end{equation}
\noindent where $\tilde{d_i}=d_i  c_n {\rm ran}(\zeta) g(l_i,b_i,d_i)$ is the perturbed distance of the $i$-th star, $g(l_i,b_i,d_i)$ accounts for the angular position of the star in the sky, ${\rm ran}(\zeta) $ is a random number generator, and $c_n$ makes sure that (\ref{av_cnd}) is satisfied. 

We select eight different values of $f_{\rm in}$ in the range $[-0.2,0.2]$ (i.e. $\pm 20 \%$). Using the $V$-$W$ estimator to estimate the mean fractional distance error on the sample, we recover the true bias within one standard deviation: $f_{\rm in}-f_{V-W}=0.0162\pm0.0187$.

\subsection{Calibration construction}\label{sec:constr}
Stellar evolution models predict a steep decline of the intrinsic magnitude of BHB stars towards bluer colours (e.g. BASTI, Dartmouth isochrones) in concordance with photometry of globular clusters. Further, they predict metal-poor BHB stars to be systematically brighter than their metal-rich counterparts. The ideal calibration would encompass the main dependences with the simplest possible form: in particular, the metallicity dependence, modelled best by a polynomial for the increasing trend in metallicity, and the steep decline on the blue side, which we model by an exponential in $(g-r)_0$ colour:
\begin{equation}\label{theoMg}
M_g((g-r)_0,[Fe/H])=a_0 \exp(a_1 (g-r)_0)+p(\fe)
\end{equation}
\noindent where $p(\fe) \in \mathbb{R}_n[\fe]$. Parameters of (\ref{theoMg}) are empirically adjusted to mimic both the trends seen in metal-poor clusters and the ones predicted by the isochrones: among the ones available the BASTI isochrones drive our calibration towards a better performance on field stars w.r.t. for example the Dartmouth ones (see \S \ref{sec:BASTIvsDart}). One must account for photometric and reddening uncertainties though, which make it likely that a star with assigned blue-hook colours will conceivably be still relatively luminous: thus the proposed calibration shall bend a bit less steeply towards the blue hook than the isochrones.

Our calibration  will not be a fit to the data, but rather it will be validated by data. The crucial difference is that a priori we do not know if a data sample is perfectly unbiased and how eventual systematics can bias the parameters of (\ref{theoMg}). On the other hand a validation method can rely on the limited information content, which proves to be systematics-free. In the BHB stars sample of \cite{x11} for example, \cite{P2} flag a peculiar kinematics associated with metal-poor high $c_\gamma$ stars: these stars affect the correlation between the $V$ component of velocity and the angle term $\sin(l)\sin(b)\cos(b)$, but have negligible effect on the relationship between $W$ component of velocity and the same angle term arising from the rest of the sample.

\section{Sample selection}

SDSS and its sequel Sloan Extension for Galactic Understanding and Exploration (SEGUE) obtained an unprecedented photometric and spectroscopic mapping of the Milky Way galaxy, allowing the study of nearby dwarf populations as well as giant star populations up to 100 kpc from the Sun \citep[][]{yan09}. The SEGUE Stellar Parameter Pipeline provides line index measurements for the spectra and the three primary stellar parameters: effective temperature ($T_{\rm eff}$), surface gravity ($\log g$) and metallicity ($\fe$) \citep[cfr.][]{lee08,lee08b,all08}. Alternative methods have been developed to estimate the above parameters: we will adopt the estimates produced by the method of \cite{wil99} as the latter was particularly designed for hot stars. 

We build a {\bf photometrically selected sample} of BHB stars drawn from the 9th Data Release \cite[DR9,][]{ahn12}: we query for BHB candidates and filter in colour (dereddened) and stellar parameters in order to minimize the risk of contamination from other luminosity types according to the warnings laid out in \cite{yan00} and \cite{sir04}. Specifically our selection criteria are

\begin{equation}\label{FScuts}
\left\{ \begin{array}{l}
g<18\\
2<\log(g)<3.5\\
0.8<(u-g)_0<1.4\\
-0.4<(g-r)_0<0 \\
7250<T_{\rm eff}/{\rm K}<9700\\
(\alpha,\delta) \notin {\rm Sgr}
\end{array}\right.
\end{equation} 

\noindent where the last criterion removes $786$ stars or respectively $30\%$ of the sample within a polygon on the sky that encloses Sagittarius, the coordinates of which were kindly provided to us by the authors of \cite{de1} and are reported in \cite{P2}.
These yield a sample of $\sim 2600$ stars. A detailed discussion on relaxing these criteria can be found in \cite{P2}. 

We also consider the {\bf spectroscopically selected sample} compiled by \cite{x11} and kindly provided to us by Xiangxiang Xue on request: this consists of $\sim 4000$ objects drawn from SDSS DR8 \citep[][]{eis11} that pass both the colour cuts of \cite{yan00} and the two Balmer-line profile cuts already described in \cite{x08}, but here slightly relaxed \citep[see][]{x11}. We retrieve these objects in DR9 and adopt the latter values for both their physical parameters and their astrometry.

As both samples cover a distance range out to $\sim 50 \kpc$, the error in proper motion measurements for single stars is approximately as large as the true signal from solar reflex motion and velocity dispersion. However, by large number statistics the proper motions still bear valuable information. While the statistical uncertainty simply sets our formal error bars, we have to be concerned with the systematic terms that are roughly a factor of $20$ smaller than the noise \citep[see \S 4 of][]{P2}. To cope with the systematics, we will use the proper-motion correction derived by \cite{sch12} on the \cite{schn10} quasar sample in Galactic angles $(l, b)$ and note that this correction shrinks the statistically predicted mean fractional distance error by $0.02$ (see \S 3.1.2).

Hereafter, we shall refer to the photometric sample selected via ({\ref{FScuts}) as FS12 and to the spectroscopic one as X11. We will assess the distance calibrations on these two samples, though we will mainly rely on X11 as it allows for better statistics.

\section{Distance calibration for the BHB}

The end product of the study outlined in \S \ref{sec:constr} is:
\begin{eqnarray}\label{Mg}
M_g ((g-r)_0,\fe) &=&  0.0075 \, \exp(-14.0 {(g-r)_0}) +\\
& & + 0.04\left(\fe + 3.5\right)^2+ 0.25 ,  \nonumber
\end{eqnarray}
\noindent which we show in Fig. \ref{bhbmag} for two metallicities: $\fe=-2.25$ (thick blue line) and $\fe=-1.25$ (thick red line). 
Also shown in Fig. \ref{bhbmag} are the BASTI isochrones \citep[][]{pie04,pie06}.
The discrepancy between the constant-luminosity approximation (horizontal light blue line in Fig. \ref{bhbmag}) and both the isochrones and our calibration exceeds the frequently adopted uncertainty band of $\varepsilon_{M_g} \sim \pm0.18$ (light blue dotted lines in Fig. \ref{bhbmag})\footnote{An uncertainty of $\pm 0.18$ is commonly associated with the calibration of the horizontal branch, due to neglect of the temperature dependence \citep[][]{sir04}, or ($\pm 0.15,\, 0.2$) associated with the constant absolute magnitude approximation \citep[e.g.][]{de1,dep10}}. 

\begin{figure}
\epsfig{file=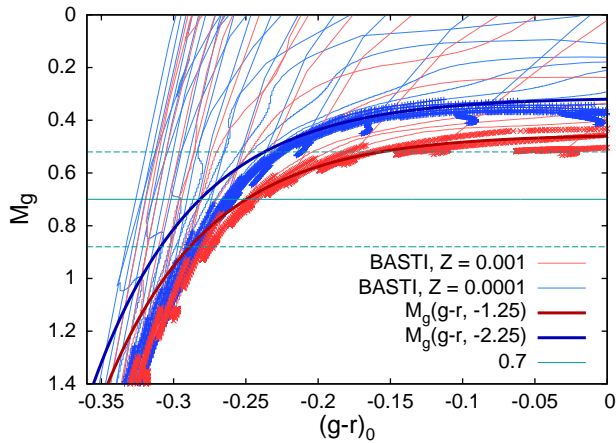,angle=-90,width=85mm}
\caption{Luminosity as a function of colour for the Blue Horizontal Branch at different metallicities. Thin lines represent the predictions from the BASTI isochrones  \citep[][]{pie04,pie06} as do the dots that are associated with a mass spacing of $3 \cdot 10^{-6} M_\odot$. Thick lines depict our colour and metallicity dependent approximation for $M_g$. Blue lines refers to metal-poor stars ($\fe \in ]-3,-2]$) and red lines to metal-rich stars ($\fe \in ]-2,0]$). The light blue line depicts the popular approximation of constant $M_g=0.7$ and light blue dotted lines represent an uncertainty of $\pm 0.18$.}
\label{bhbmag}
\end{figure}

Equation (\ref{Mg}) does not depend on the $u-g$ colour as in our tests we could not detect any significant systematics for it once the $g-r$ and $\fe$ dependences had been taken into account.
The ideal calibration would account for more detailed chemical dependences like alpha enhancement though and particularly the abundance of Helium, which changes the absolute magnitude of a star. However, this information is generally not available for BHB stars as even though they are very hot their Helium abundance can be very low due to depletion mechanisms \citep[e.g.][]{fau67}.

\subsection{Validation on field stars}

We validate our calibration on our two samples of halo field BHB stars: on both we obtain a satisfactory accuracy:
$$
f_{\rm FS12}=0.01 \pm 0.04 \textrm{\quad and \quad} f_{\rm X11}=0.02 \pm 0.03.
$$
Given that the fractional average distance error from (\ref{Mg}) is consistent with zero within the statistical uncertainties, our calibration can be regarded as the first direct measurement of the magnitude of field BHB stars in SDSS-SEGUE.

Use of the proper motion\footnote{We checked that the line-of-sight velocity correction given in the same paper has virtually no effect on our distance estimates.} correction of \cite{sch12} reduced the average systematic distance error from $f=-0.03 \pm 0.04$ to $f=0.01\pm 0.04$ for the FS12 sample and from $f=-0.04 \pm 0.03$ to $f=0.02\pm 0.03$ for X11 using our calibration (\ref{Mg}) for the distance estimate. Hence, the necessary correction affects the average distance calibration by about $5 \%$ and hardly affects the relative distances between single bins.

In general the statistics are robust against outliers in the sense that when we remove stars with extreme proper-motion (i.e. tangential velocities above $900 \, \kms$: less than $3 \%$ in FS12 and than $6 \%$ in X11), $f$ does not change for FS12 and changes by less than 0.01 for X11. The inclusion of Sgr makes very little difference ($\Delta f \sim 0.02$) to our estimates. Indeed, adding the Sagittarius region to the samples,  we find $f=0.04 \pm 0.03$ on X11 and a similar difference on FS12: $f=0.02 \pm 0.04$.

\section{Discussion of other magnitude/distance assignments}

Using the same method that allowed us to validate our calibration against field halo BHB stars (see \S \ref{sec:two}), we can directly assess other distance calibrations.

\subsection{BASTI versus Dartmouth isochrones}\label{sec:BASTIvsDart}

In Fig. \ref{DB} we plot the discrepancy between the absolute magnitude predicted by the Dartmouth isochrones and our calibration as a function of colour for four synthetic populations of $\sim 2700$ HB stars, each with different metallicity. The synthetic populations were generated with the Dartmouth web tool (see http://stellar.dartmouth.edu/models/shb.html) and input parameters $([\alpha/Fe],\max M, \langle M \rangle, \sigma_M)=(0.4,4.0, 0.5,0.1)$, where $M$ is mass in solar masses. In the colour range $g-r \in [-0.25,0]$ (where our sample stars are concentrated) the offset is systematic and increases with metallicity: $\Delta M_g \sim 0.1$ for metal-poor objects and $\Delta M_g \sim 0.3-0.4$ for metal-rich objects.\footnote{Varying $[\alpha/Fe]$ by $\pm 0.2$ does not alter our conclusions.} Given that we achieve a fractional average distance error consistent with zero well within $1 \sigma$ on our sample of field stars, estimating the stars' absolute magnitudes with the Dartmouth isochrones corresponds to underestimating distances by up to $4 \, \%$ for metal-poor stars and by up to $17 \, \%$  for metal-poor ones (see eq. \ref{convert} below). 

In conclusion, the close agreement between our formula and the BASTI isochrones (see Fig. \ref{bhbmag}) implies that at least in the $ugriz$-system, they are superior to the Dartmouth isochrones.

\begin{figure}
\epsfig{file=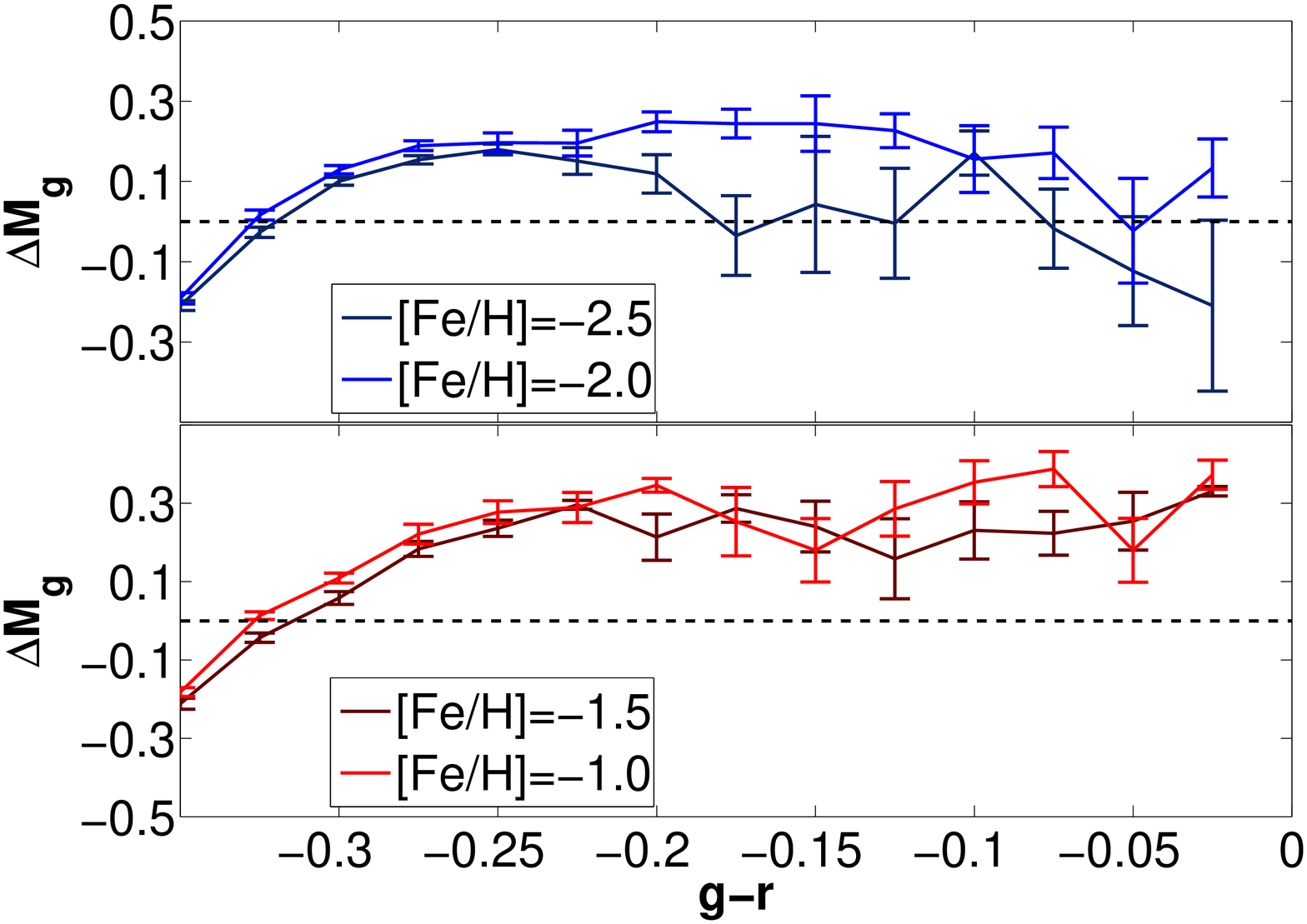,angle=0,width=85mm}
\caption{Discrepancy between the absolute magnitude predicted by the Dartmouth Isochrones and the one estimated via equation \ref{Mg} as a function of colour for four synthetic populations of $\sim 2700$ HB stars each with different metallicity. Upper panel: metal poor stars; bottom panel: metal rich stars. Error bars show the standard error on $\Delta M_g$ in each colour bin.}
\label{DB}
\end{figure}

\subsection{Sirko et al. 2004}
\cite{sir04} estimated the absolute magnitudes of BHB stars as functions of colour for two metallicities (${\rm[M/H]} =-1$ and ${\rm[M/H]} =-2$) by integrating the Kurucz model atmospheres\footnote{see http://www.kurucz.harvard.edu.} and then finding the point on the theoretical track that was closest to the observed star in colour space. They validated their calibration with two globular clusters: NGC 2419 and Pal 5. With literature distance moduli for the clusters they found that the isochrones were too bright by $0.13 \mag$ for NGC 2419 and and too faint by $0.2 \mag$ for Pal 5. They considered that these offsets could be accounted for by errors in the adopted distance moduli.

When we compare our calibration with theirs, it turns out that their relation is significantly fainter than ours: $f_{\rm Sirko}=-0.07 \pm 0.04$ on X11. Hence \cite{sir04} predict shorter distances than our formula, which on our data produces a fractional average distance error consistent with zero well within 1 $\sigma$ (see \S 3.1). Further application of the calibration of \cite{sir04} (from their Table 2) requires several approximations: one needs to bin in effective temperature, surface gravity, colour and metallicity. Consequently the calibration is undefined for more than half of the stars in the sample.\footnote{To avoid binning the data, we compared the two formulas at fixed metallicity. We use $\fe$ while \cite{sir04} use $[{\rm M/H}]$, but this has no relevance in our discussion. When equating $\fe$ and $[{\rm M/H}]$ and defining $\Delta M_g=M_g ({\rm Sirko})-M_g({\rm FS12})$, we find $\Delta M_g \in [0.07,0.35]$ for metal-rich stars and $\Delta M_g \in [0.15,0.41]$ for metal-poor stars. When we assume $[{\rm M/H}]=\fe+0.3$, which roughly corresponds to the \cite{sal93} approximation $[{\rm M/H}]=\fe+\log(0.638 \exp([\alpha/{\rm Fe}])+0.362)$ with $[\alpha/{\rm Fe}] \sim 0.4$, the estimates above are unchanged.  In conclusion, given their fainter magnitude, \cite{sir04} calibration will produce shorter distances w.r.t. (\ref{Mg}) by a $\Delta f$ of at least $-0.03$ and up to $-0.17$. Thus, it is reasonable to expect that on our samples their calibration would produce a distance error in the range $f \in [-0.15, -0.02]$, regardless of how data are binned.}

Our calibration is an analytic formula and not a complicated fitting process to the theoretical models. Since it produces significantly smaller distance errors than the calibration of \cite{sir04}, it is not only more convenient, but also more accurate.

\subsection{Xue et al. 2011}

 The catalogue of BHB stars provided by \cite{x11} includes distances estimated with a method very similar to \cite{sir04}: i.e. the authors determine the point on the theoretical tracks that is closest to the observed star in $u-g$, $g-r$ space and from the former estimate $M_g$. On X11, i.e. the catalogue of \cite{x11} with the Sgr region removed, their distances appear to be weakly biased towards distance underestimates: $f=-0.05 \pm 0.03$.

\begin{figure}
\epsfig{file=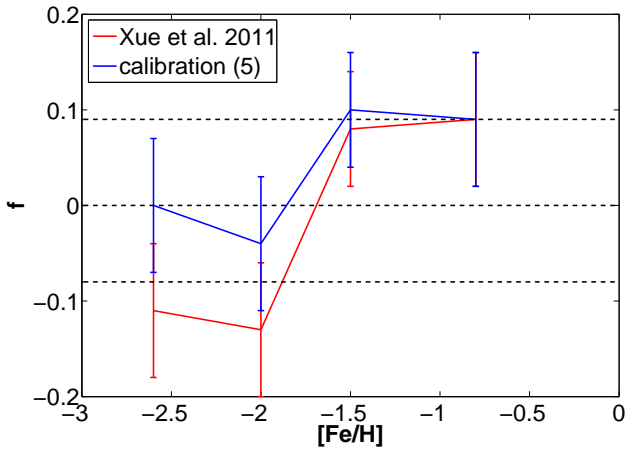,angle=0,width=85mm}
\caption{Average fractional distance error as function of metallicity on X11 sample, using both the calibration of Xue et al. (2011, red line) and equation \ref{Mg} (blue line). }
\label{x11_dist}
\end{figure}

Marginalizing over metallicity, the fit to the colour dependence is comparable between their calibration and ours, within the formal errors. However, our formula (\ref{Mg}) proves to be significantly more accurate at low metallicity when we marginalize over colour: Fig. \ref{x11_dist} shows the different trend in metallicity of $f$ for the two distance calibrations on X11 sample. In the range $\fe \in [-3,-1.8]$ the calibration of \cite{x11} is associated with an average fractional distance error of $f \sim -0.12\pm 0.04$, while equation (\ref{Mg}) yields $f \sim -0.02 \pm 0.04$.

\subsection{Deason et al. 2011b}

\cite{de11_416} propose a calibration of BHB stars derived from 10 star clusters published in \cite{an08} covering a range in metallicity of $\fe \in [-2.3,-1.3]$. They detect a colour, but no obvious metallicity dependence: given the considerable uncertainties in the adopted distance moduli and reddenings \citep[$0.20\pm 0.04 \mag$ or $\sim 10-15 \%$, e.g.][]{gra97} and the systematic differences between clusters and field halo stars though, this result is not enough to rule out a detectable metallicity dependence on field BHB stars. The calibration of \cite{de11_416} is systematically fainter for metal-poor stars and brighter for metal-rich stars than ours and leads to a worse performance on the FS12 sample.

On the X11 sample we estimate a global fractional average distance error of $f=-0.02 \pm 0.03$ using their calibration ($f=0.02 \pm 0.04$ for metal-rich stars and $f=-0.08\pm 0.06$ for metal-poor stars), while on FS12 \cite{de11_416} gives $f=-0.03 \pm 0.04$  ($f=0.01 \pm 0.05$ for metal-rich stars and $f=-0.05\pm 0.06$ for metal-poor stars). When investigating the detailed morphology of the halo, this metallicity-related bias is both significant in magnitude and relevant for the research goal. E.g. substructures can be blurred and distorted by several kpc. In particular, the distances for metal-poor stars are underestimated, while that for the metal-rich ones are slightly overestimated: this gives rise to a kinematic bias acting in opposite directions.

\subsection{Systematics of the approximation $M_g=0.7$}\label{sec:seven}

Fig. \ref{dd07_ddR} shows the discrepancy in heliocentric distance between the approximation $M_g=0.7$ and our calibration on the two samples FS12 and X11: the constant absolute magnitude approximation tends to underestimate stars' distances by several kpc, the discrepancy is most severe for the most metal-poor objects. A direct assessment of the constant absolute magnitude approximation via the method in \S \ref{sec:two}  measures a global distance underestimate of $f = -0.13 \pm 0.03$ on X11 sample and of $f =  -0.11 \pm 0.04$ on FS12 sample, the error being worse for metal-poor objects. In a kinematic analysis for example, this systematics can distort the velocity distribution of the sample and the fact that the effect is stronger on metal-poor objects can introduce a false gradient in the kinematics of sub-samples of stars with different metallicities.

\begin{figure}
\epsfig{file=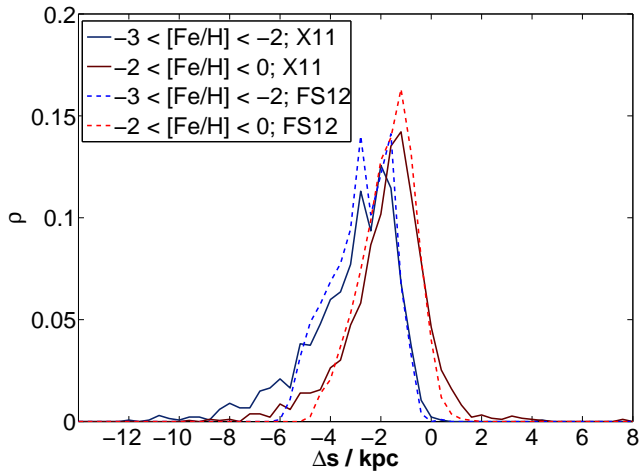,angle=0,width=85mm}
\caption{Density distribution ($\rho$) of the difference in distance calibrations for BHB stars from SDSS DR9: $\Delta s=s(M_g=0.7)-s(M_g((g-r)_0,\fe))$. Solid lines refer to X11 sample, dotted ones to FS12. The assumption of a constant absolute magnitude underestimates stellar distances w.r.t. the colour and metallicity dependent estimate for $M_g$ given by equation \ref{Mg}. The effect is more pronounced for the most metal poor objects (blue lines). }
\label{dd07_ddR}
\end{figure}

We show that the method presented in \S \ref{sec:two} is sensitive enough to detect the metallicity and colour dependence of absolute magnitudes: for this purpose we bin both the spectroscopic and the photometric samples separately in colour and metallicity to measure the expected dependences on each quantity while marginalising over the other.

\subsubsection{Colour dependence}

We divide X11 and FS12 into metal-poor, $\fe \in [-3,-2]$, and metal-rich stars, $\fe \in [-2,0]$: each sub-sample is then split in colour bins, the sizes of which are adjusted to provide enough statistics for each colour band, and therefore meaningful error bars ($\ge 400$ stars per bin).

In  Fig. \ref{f_gmr_MP_MR} we plot $f$ against $(g-r)_0$: the red lines  in the upper and lower panel are associated with the metal-rich stars in the X11 and FS12 samples respectively. The average relative distance error $f$ oscillates with colour: the trend is most pronounced in the X11 sample. This behaviour reflects the dependence of their absolute magnitudes on colour. It is not matched by the metal-poor part of either of the samples: the blue lines in Fig. \ref{f_gmr_MP_MR} show that both the photometrically selected metal-poor stars (lower panel)  and the ones spectroscopically selected (upper panel) exhibit an increase of $f$ with colour. The discrepancy in the colour-dependence at different metallicity is the footprint of the metallicity dependence of $M_g$. We remark that the metal-poor stars in the X11 sample (Fig. \ref{f_gmr_MP_MR}, upper panel, blue line) exhibit exactly the behaviour predicted in the previous section and are indeed associated with the poorest fit by a constant absolute magnitude. Also, it is at these metallicities that our distance calibration differs most from the one associated with $M_g=0.7$: Fig. \ref{dd07_ddR} clearly shows that our approximation precisely corrects for the underestimate in distance caused by the assumption of a too faint constant magnitude.

\begin{figure}
\epsfig{file=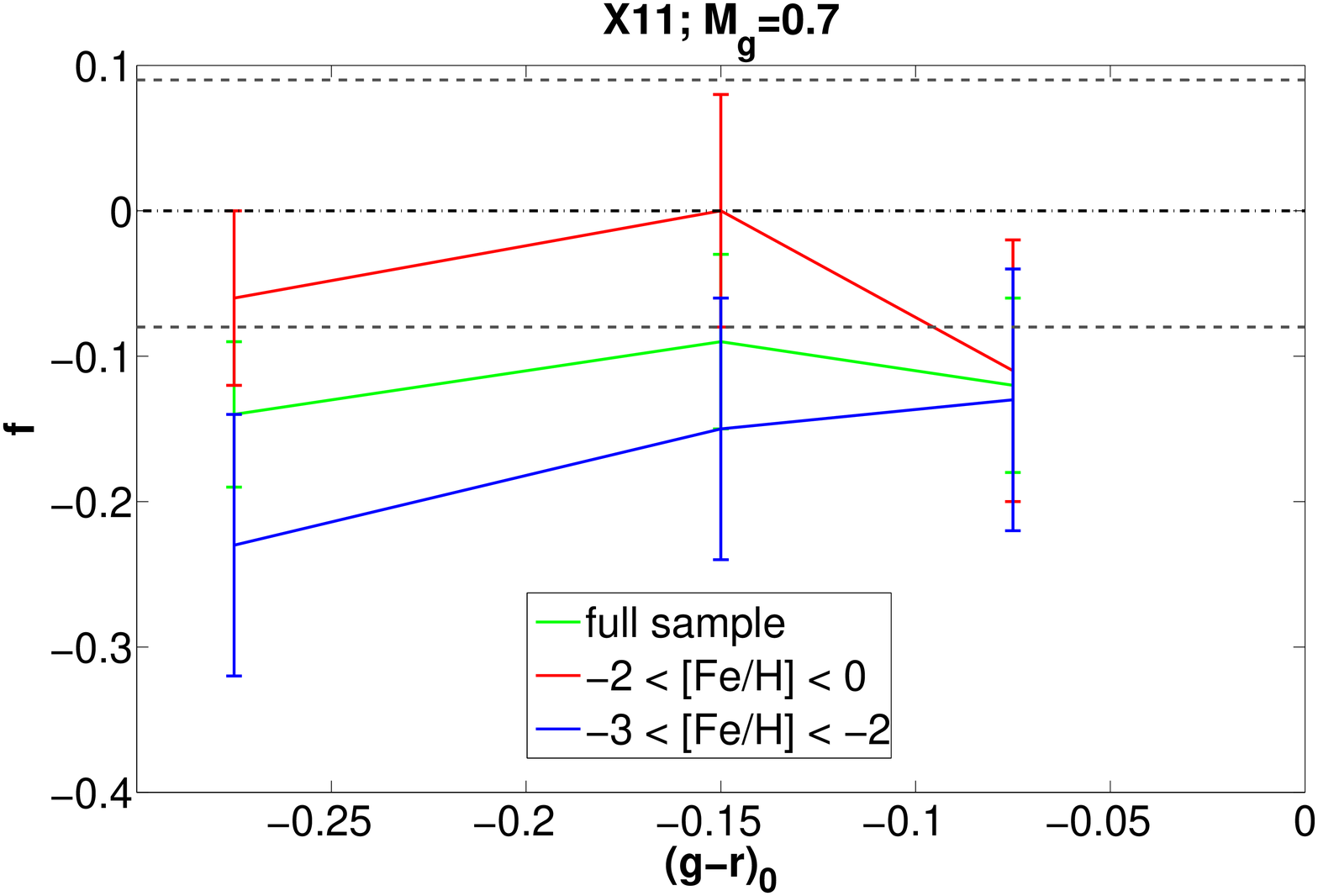,angle=0,width=85mm}
\epsfig{file=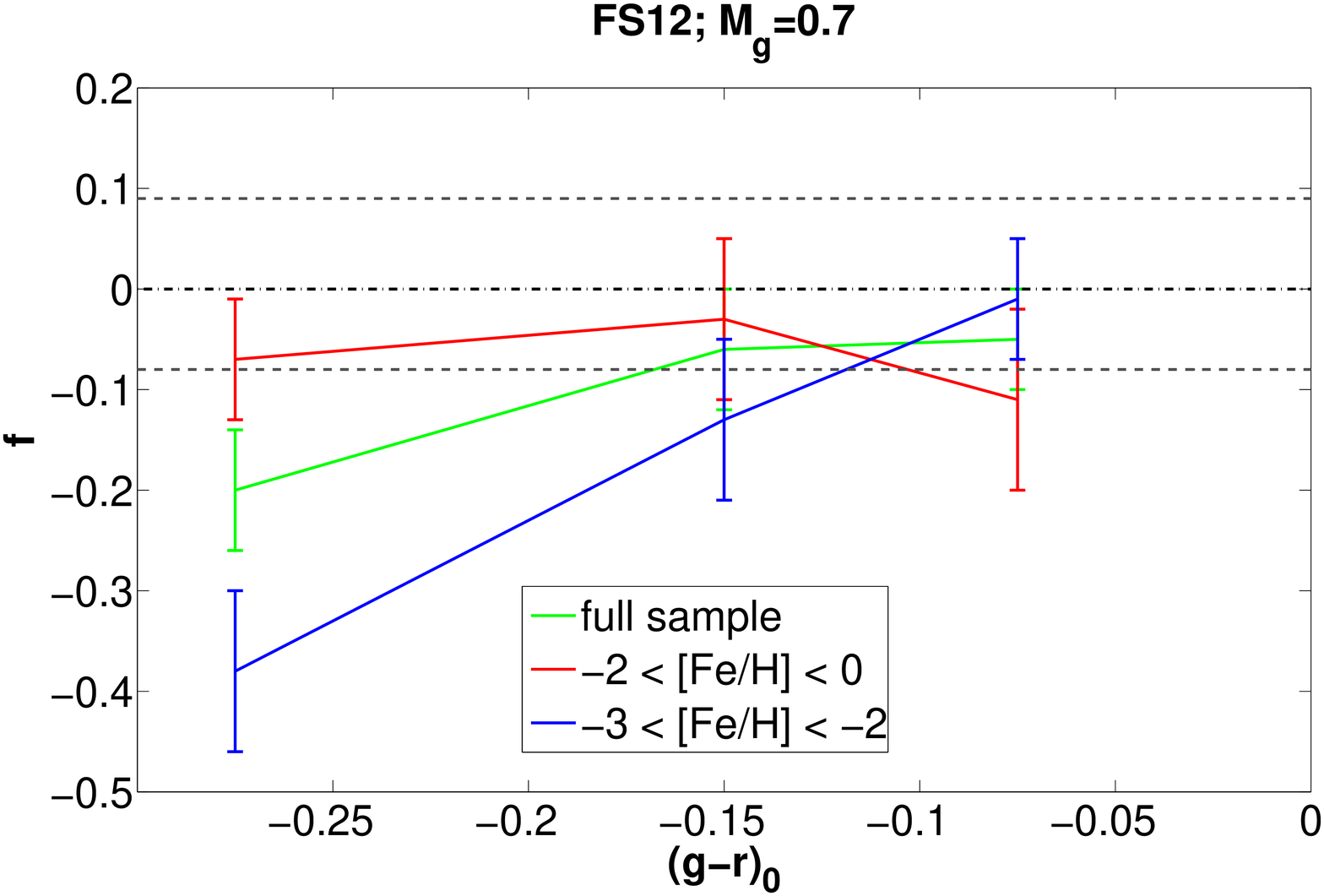,angle=0,width=85mm} 
\caption{Fractional average distance error as a function of colour at different metallicities under the assumption that $M_g=0.7$ of the spectroscopic sample X11 (upper panel) and the photometric sample FS12 (lower panel). The green lines are the full sample, while the red lines are associated with metal-rich stars and the blue ones with the most metal-poor objects. The actual horizontal coordinate has been offset for each line for the error bars to be distinguishable. Every data point is associated with the centre of a colour bin: its size varies in order for every bin to contain a comparable amount of stars and therefore produce meaningful statistics. Grey dotted lines represent an absolute magnitude uncertainty of $\pm 0.18$ \citep[e.g.][]{sir04}.}
\label{f_gmr_MP_MR}
\end{figure}

To assess the significance of these dependences w.r.t. the traditional level of uncertainty associated with the absolute magnitude of BHB stars \citep[e.g. $\varepsilon_{M_g}=0.18$,][]{sir04,dep10,de1,new09}  we need to relate the fractional distance error to the effective error in absolute magnitude, $\varepsilon_{M_g}=M_g^\prime-M_g$. Given that

\begin{equation}
M_g=g_0-5 \log_{10}\left(\frac{s}{0.01 \kpc}\right),
\end{equation}

\noindent and $s^\prime=(1+f)s$, where $s$ is the real heliocentric distance, then 

\begin{equation}
M_g^\prime=g_0-5 \log_{10}\left(\frac{s^\prime}{0.01 \kpc}\right) = M_g - 5 \log_{10}(1+f)
\end{equation}\label{convert}

\noindent so: $\varepsilon_{M_g}=-5 \log_{10} (1+f)$ . An absolute-magnitude uncertainty of $\pm 0.18$ will therefore correspond to $f \in [-0.08,0.09]$. Fig. \ref{f_gmr_MP_MR} shows that the fractional distance error trends as a function of colour for different metallicities, significantly exceed this band (grey dotted lines). This implies that the effects of approximating the absolute magnitude of BHB stars with $M_g=0.7$ go beyond the generally acknowledged uncertainty of this approximation.

Further, the effects of distance misestimates superpose ‪as they ought to‬: in Fig. \ref{f_gmr_MP_MR}, the green lines that represent the full samples are consistent with being averages of the blue and red lines associated with the metal-poor and metal-rich sub-samples respectively. Also, both samples confirm that the blue end experiences the worst fitting by a constant absolute-magnitude distance scale. The dramatic drop in the metal-rich side of the photometric sample (Fig. \ref{f_gmr_MP_MR}, bottom panel, red line) is however beyond predictable behavior and we therefore flag this end as possibly affected by biases outside our control. Unfortunately, the low statistics associated with this sample do not allow a more detailed investigation of the problem.

Fig. \ref{f_gmr_all} shows the dependence of $f$ on $(g-r)_0$ colour on both samples for our calibration and the constant absolute magnitude approximation. The performance of (\ref{Mg}) on the spectroscopic sample X11 (blue solid line) is almost perfectly satisfactory, while FS12 (blue dotted line) appears slightly problematic at the blue end as expected. We cautiously avoid extensive comment on the photometric sample as regards colour dependence due to the low statistics available for the blue end and the large uncertainties in the stellar parameters at that end.

\subsubsection{Metallicity dependence}
Fig. \ref{f_fe_all} shows the dependence of $f$ on metallicity for both distance calibrations. $|f|$ increases towards underestimates in distance (negative $f$) with decreasing metallicity in both samples (Fig. \ref{f_fe_all}). The discrepancy between the two distance calibrations is also affected by metallicity: still in Fig. \ref{f_fe_all}, we see that the offset between the blue lines (associated with the colour and metallicity dependent calibration) and the red lines (constant absolute magnitude) is largest at the metal-poor end for both the FS12 and X11 samples. 

The fact that the most metal-poor objects suffer the largest inaccuracy in distance (already visible in Fig. \ref{f_gmr_MP_MR}, upper panel), and experience the largest discrepancy between the two distance calibrations explains why in Fig. \ref{f_gmr_all} the offset between the two distance calibrations is significantly larger for the X11 sample w.r.t. the FS12 one: the former sample is more metal-poor than the latter and the different mixture of metallicities translates into a larger difference between $M_g=0.7$ and our formula (\ref{Mg}). 

Finally, we note that when we adopt the colour and metallicity-dependent distance calibration, $f$ lies within the traditional uncertainty band associated with $\varepsilon_{M_g}=\pm 0.18$ (corresponding to $f=-0.08$ and $f=0.09$ respectively) for both the spectroscopic and photometric sample and both w.r.t. colour and metallicity. In particular we tested the hypothesis that $f$ is uncorrelated with metallicity when (\ref{Mg}) is adopted: a chi-squared test confirms that the discrepancy between the data and a constant trend is not enough to reject the null-hypothesis, 10-20 $\%$ significance level (s.l.) for X11 and 30-50 $\%$ s.l. for FS12.

On the other hand the same hypothesis is clearly rejected for the trends associated with $M_g=0.7$ implying a statistically significant correlation between $f$ and $\fe$: at the $1 \%$ s.l. for X11 and at the $5-10 \%$ s.l. for FS12.

\begin{figure}
\epsfig{file=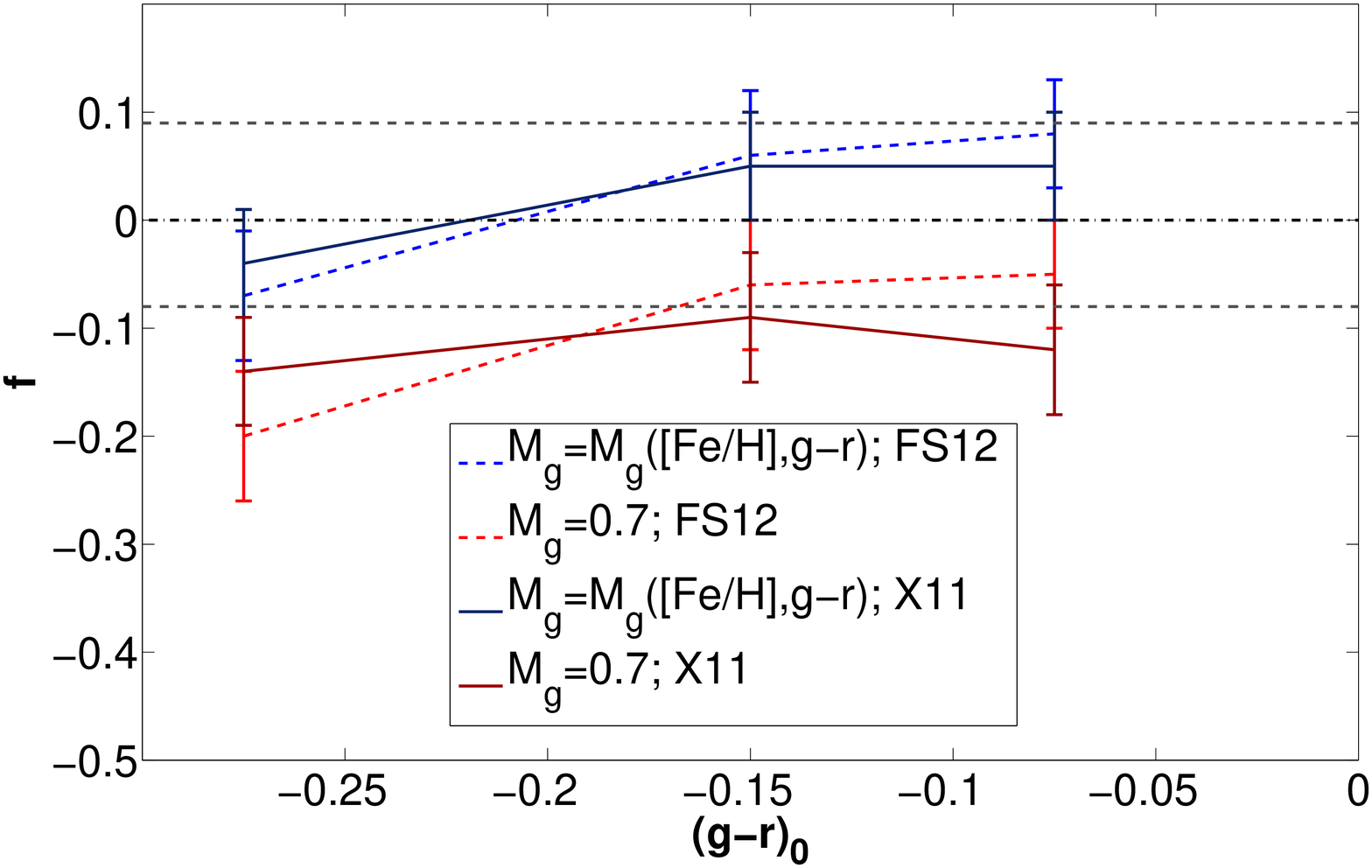,angle=0,width=85mm}
\caption{Fractional average distance error as a function of colour for BHB stars from SDSS DR9. Blue lines refer to the colour and metallicity dependent distance calibration (\ref{Mg}), red lines depict the constant $M_g=0.7$. Solid lines and lighter colours are associated with the spectroscopic sample X11, dotted lines and darker colours with the photometric sample FS12. The discrepancy between the two distance calibrations is larger for X11 due to the sample being more metal-poor than FS12 (see text). Grey dotted lines represent the quoted absolute magnitude uncertainty of $\pm 0.18$.}
\label{f_gmr_all}
\end{figure}

\begin{figure}
\epsfig{file=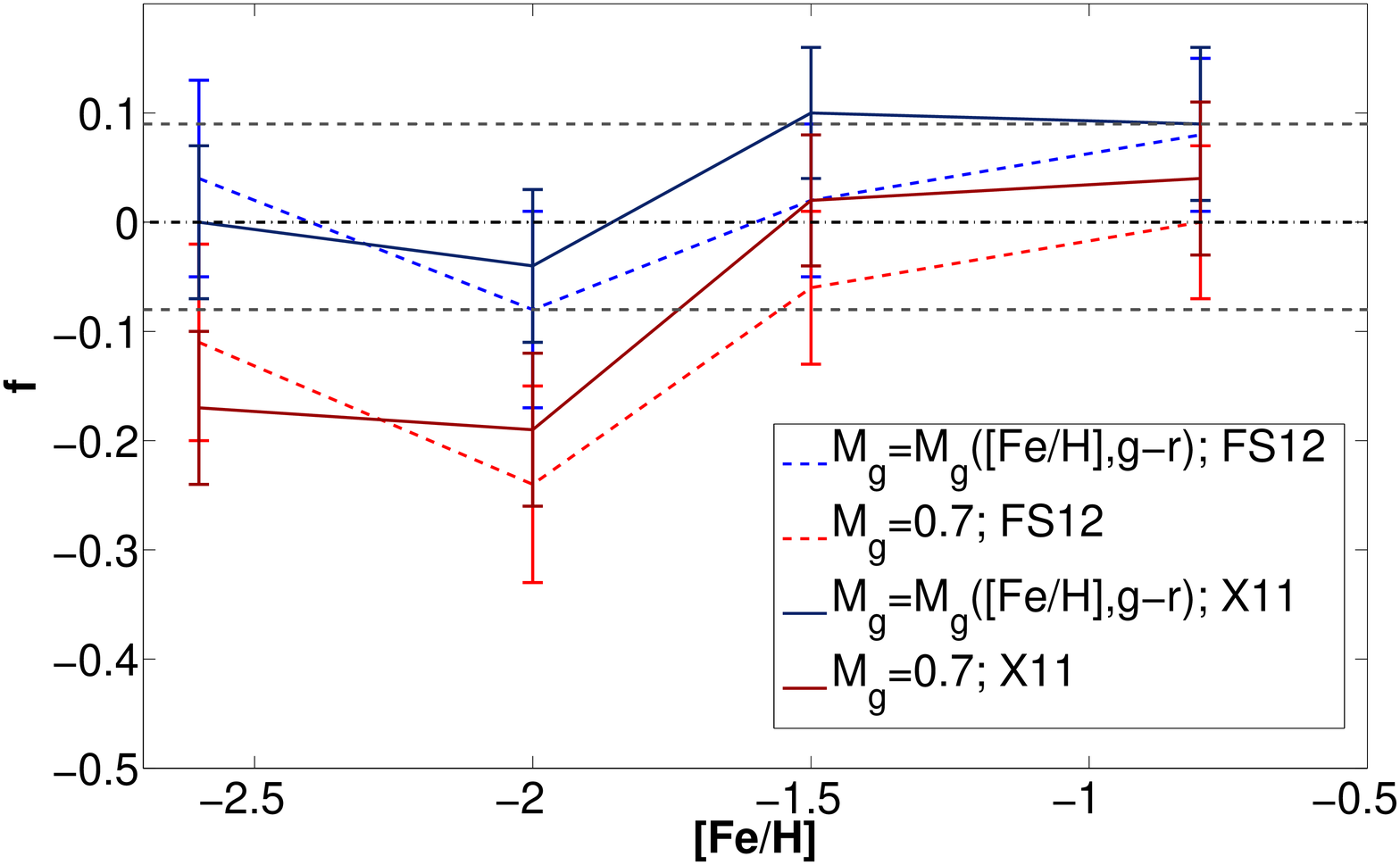,angle=0,width=85mm}
\caption{Same as Fig. \ref{f_gmr_all}, but here we plot the fractional average distance error as a function of metallicity for the two samples.}
\label{f_fe_all}
\end{figure}

\subsection{Improvement w.r.t. previous calibrations}

In Fig. \ref{cal_comp} we show a comparison between the calibrations considered above: the plot is an extension of Fig. \ref{bhbmag} in the sense that it now shows the trend of $M_g$ with $(g-r)_0$ colour at two different metallicities also for the calibrations of \cite{sir04,x11} and \cite{de11_416}.  We do not re-plot the isochrones to avoid crowding in the figure.

\begin{figure}
\epsfig{file=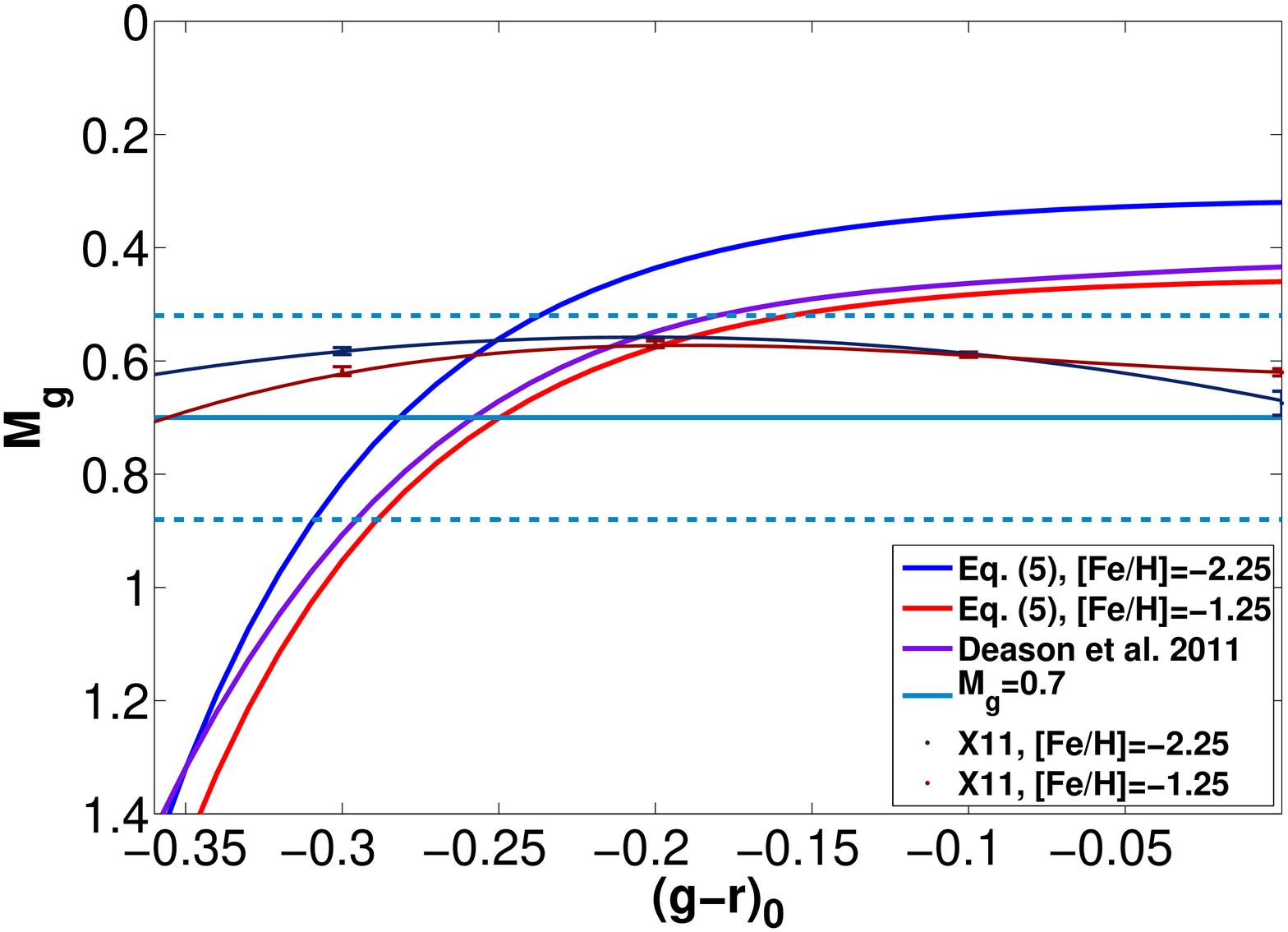,angle=0,width=85mm}
\caption{Same as for Fig. \ref{bhbmag} but including the calibrations of Sirko et al. (2004), Xue et al. (2011) and Deason et al. (2011b). The calibration of Deason et al. (2011b) is independent of metallicity and hence shown only once.}
\label{cal_comp}
\end{figure}

We find that our calibration is the only one among the ones listed above not to suffer the ``falling sky" pathology of SBA12 (see \S \ref{sec:two}), with the other calibrations showing the most severe effects on the metal-poor sub-sample, as expected from the previous assessment. In Fig. \ref{fallingsky} we show the average correlation between $W$ and $\sin l \sin b \cos b$ for the metal-poor objects in the X11 sample for our calibration, the one of \cite{sir04, x11} and for the constant magnitude approximation. The performance of the calibration of \cite{de11_416} is almost identical to the ones of \cite{sir04} and \cite{x11} and we avoid plotting it in order not to crowd the figure. We note that the offset from zero in the mean $W$ disappears accounting for a $vlos$ aberration of $\sim 10-15 \kms$. While this is not a proof that SEGUE $\vlos$ suffer from a systematic offset, the coincidence requires further investigation: we reserve this for future work.

\begin{figure}
\epsfig{file=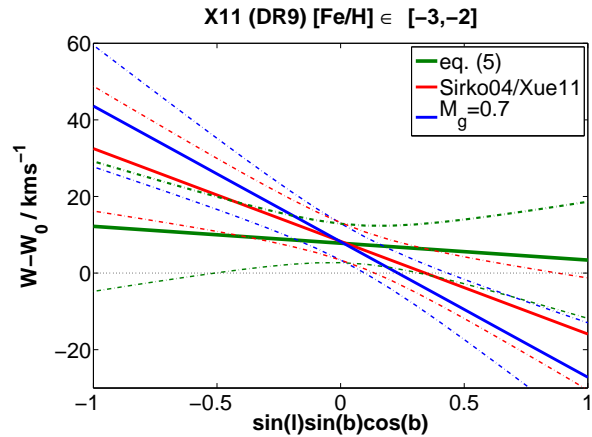,angle=0,width=85mm}
\caption{Mean correlation between the $W$ component of velocity and the angular term $\sin l \sin b \cos b$ for metal-poor stars ($\fe < -2$) in the X11 sample. Solid lines are the mean trends associated with the different calibrations: green for our eq. \ref{Mg}, red for the calibration of Sirko et al. (2004), Xue et al. (2011) and blue for the constant absolute magnitude approximation. With the same colour code, dashed lines represent the $1\sigma$ contours.}
\label{fallingsky}
\end{figure}

\section{Conclusions}

We have proposed and validated a very simple analytic formula that properly describes the colour and metallicity dependence of the absolute magnitudes of BHB stars. Our calibration both provides a user-friendly tool to compute the absolute magnitude of BHB stars from observables in the $ugriz$-system and achieves the best performance on field BHB halo stars w.r.t. other available calibrations. Given the agreement between our formula and the BASTI isochrones, we conclude that these isochrones successfully predict $ugriz$ colours for the blue part of the HB. The Dartmouth isochrones are fainter than the BASTI models by $\sim 0.1-0.4 \mag$ and therefore lead to underestimate the distances of field BHB stars.

The quality of proper motions in the SDSS is sufficient to assess the average distance scale. Even though the large distances covered by the two samples imply large random errors in kinematics from proper motions, our considerations are based on averages across the sky. The linear estimator we adopted from SBA12 is therefore not affected by the inaccurate knowledge of the size of proper-motion errors, while it is mildly vulnerable to proper motion systematics. We corrected for them following \cite{sch12} and showed that this correction leads to a stretch of the distance scale of about $5 \%$.

Another issue is the unknown level of contamination by blue stragglers and main-sequence stars. This contamination will bias our distance estimates towards signalling an overestimate. Hence our formula providing formally unbiased distance estimates may still be rather on the faint side despite being brighter than the alternative calibrations. Our magnitude approximation may be too flat at the blue end of the horizontal branch, where it declines towards the blue hook. We have actively chosen this mild deviation in the light of reddening and photometric uncertainties.

The fact that the absolute magnitude of BHB stars depends on both colour and metallicity implies that any calibration that does not account for this dependence is a priori biased. We have shown that impact of a constant magnitude approximation on distances exceeds the $10 \%$ level ($f \in [-0.38,0.04]$ or equivalently $\Delta M_g \in [-0.09,1.04]$, see Fig. \ref{f_gmr_MP_MR}, \ref{f_fe_all}) and that a calibration which is dependent on colour, but not on metallicity \citep[e.g.][]{de11_416} introduces a systematics in distance (of order of $0.23 \mag$ or respectively $10 \%$) which correlates with metallicity. This effect is of particular concern in kinematics studies for it can produce a false kinematics gradient across populations with different metallicity. 

The available calibrations that depend on both colour and metallicity \citep[][]{sir04, x08,x11} require interpolation of theoretical isochrones for each magnitude estimate and they were validated on a handful clusters, which themselves have a distance uncertainty.  We showed that on field BHB stars they significantly under-perform our suggested calibration (on metal-poor stars $f$ can be as high as $0.17$) and actually fail the ``falling sky" test of SBA12.

The simple calibration formula we presented in equation \ref{Mg} of this paper proves to be more accurate than any of the above on field BHB stars, achieving an accuracy that matches the threshold of the noise determined by the sample size, being the only one that passes the ``falling sky" test of SBA12. Larger information content of the data (e.g. better metallicity estimates or a new dimension like Helium content) and better statistics would be required to improve the accuracy of our formula and assess its performance in the blue-hook region. 

In general, the method we use to assess distance calibrations can be applied to any filter set and sample of field stars in the sky. The only requirements are that the sample comprises a sufficiently large number of objects with acceptable proper motion quality, is sufficiently extended in Galactic coordinates and especially is not associated with a strong kinematic selection or is dominated by a low number of streams.

\section{Acknowledgments}
We thank James Binney for useful discussions and helpful comments on a draft of this paper. We gratefully acknowledge inputs by Heather Morrison on an early draft. ‪David Weinberg‬ greatly helped improving the communication of these results. F.F. acknowledges financial support from the Science and Technology Facility Council (UK) and from Merton College, Oxford. R.S. acknowledges financial support by NASA through Hubble Fellowship grant $HF-51291.01$ awarded by the Space Telescope Science Institute, which is operated by the Association of Universities for Research in Astronomy, Inc., for NASA, under contract NAS 5-26555. Funding for SDSS-III has been provided by the Alfred P. Sloan Foundation, the Participating Institutions, the National Science Foundation, and the U.S. Department of Energy Office of Science. The SDSS-III web site is http://www.sdss3.org/

\bibliographystyle{hapj}

\bibliography{Fermani_Schoenrich_a}

\label{lastpage}

\end{document}